\documentclass[%
 reprint,
 amsmath,amssymb,
 aps,
]{revtex4-1}

\usepackage{graphicx}
\usepackage{dcolumn}
\usepackage{bm}

\begin{document}


\title{Temporal super-resolution imaging inspired by structured illumination microscopy}

\author{Farshid Shateri}
\author{Mehdi Hosseinalizadeh}%

\author{Zahra Kavehvash}
\email{kavehvash@sharif.edu}
\affiliation{%
Sharif University of Technology, Department of Electrical engineering, Azadi Ave., Tehran, Iran.
}%

\date{\today}

\begin{abstract}
In this paper, a method for increasing the temporal resolution of a temporal imaging system has been developed. Analogously to the conventional spatial imaging systems in which resolution limit is due to the finite aperture of the lens, in a temporal imaging system, finite temporal aperture of the time lens is responsible for limited temporal resolution. Based on the method used in spatial structured illumination super-resolution microscopy in spatial optics, we have utilized time prisms, which are temporal versions of conventional prisms, to shift the illusive frequency components of the input signal to become captured by the time lens. This method would pave the way to a high-resolution temporal imaging system which has applications in the observation and study of fast and rare phenomena such as dynamics and evolution of optical rogue waves or cancer cells in the blood.
\end{abstract}

\pacs{Valid PACS appear here}
\maketitle


\section{Introduction}\label{int}
In the last few decades, based on a mathematical duality between dispersion of an optical narrow-band pulse and diffraction of a paraxial beam, a field of research known as "temporal optics" has drawn great attention \cite{kolner94}. Based on the mentioned duality, many theories and applications in well-established science of optics and imaging now have a counterpart in temporal optics and temporal imaging systems (TIS). For example, time prism (TP) and time lens (TL) are two of the most important elements in temporal optics which have quite well-known analogs in spatial optics \cite{salem,howe}. Even further, the temporal equivalent of some of the most familiar optical ideas, theories, and laws such as Snell's law, Talbot effect, cloaking, image formation, and optical Fourier transform, has also been theorized and implemented \cite{talbotazana}, \cite{goodman}, \cite{moticloak}, \cite{Pengyu}, \cite{Zhao}. One of the most useful applications of optics is imaging technology which is ubiquitous in industry, daily life, and military. Speaking of imaging technology, one of its most important fundamental characteristics is the resolution, both spatial and temporal.\\
Spatial resolution limitation was first analyzed by Ernest Abbe in 1873 where he found a fundamental limit on the resolution of an imaging system known as "diffraction limit". He realized that this limit stems from two reasons: first, the wavelength of the light and second, the numerical aperture of the optical instrument. Even though this limit was considered as an inevitable and unsolvable restriction of imaging systems, in recent decades, super-resolution (SR) techniques have come into existence to alleviate this problem.  Among them the most famous is the structured illumination microscopy (SIM) which leads to an effective increase of the detection pass-band compared to head-on low numerical aperture (NA) illumination. Structured illumination-based super-resolution (SISR) is one of the most approached and well-known optical super-resolution techniques. This technique increases the resolution of an imaging system beyond the diffraction limit by retrieving the missed high-frequency components of the object using a diffraction grating \cite{Wilde16,Flu}.\\
As the finite aperture of a spatial lens is responsible for limited resolution of the system, in temporal imaging systems, the finite temporal aperture of the TL causes a limited resolution of a temporal optical imaging system. Temporal optical systems have immense applications in the study of ultrafast phenomena, optical communication, spectroscopy, and etc \cite{salem}. Therefore, as huge amount of efforts is made to overcome the diffraction limit in the spatial imaging systems, temporal resolution limit which is named here as dispersion-limit due to its spatial dual, must also be handled.\\
Temporal super-resolution has been addressed in a recent work by Yaron et al \cite{yaron} based on localization microscopy algorithm. Still, this work suffers from the fact that an \textit{a priori} knowledge of the input pulse is need, therefore, their scheme is not suitable for statistically rare pulses of which we do not have any information. Time-lens array imaging has also been proposed in another work \cite{zhan} to improve the temporal resolution. Yet, as will be explained in section \ref{Theo}, adding lenses in the time-lens array would rise another problem. In spatial optics, although utilizing a large lens means allowing higher angular frequency components to pass through the lens, it would lead those high frequency components to go beyond the paraxial regime. Same problem arises when using a lens-array, therefore, since paraxial beams and narrow-band pulses are theoretically similar, using multiple time-lenses in an array would violate the narrow-band assumption in temporal systems.\\
In this paper, an approach based on the spatial SISR technique is proposed in time-domain in order to enhance the resolution limit introduced by the finite aperture of the TL. In the proposed modified temporal imaging structure, two TPs (made of an electro-optic modulator) are utilized along with corresponding time-lenses for shifting the high-frequency components of the input signal and thus passing them through the limited aperture of the TL. This will help in capturing a higher range of object's temporal frequency bandwidth which in turn improves the output image resolution. Finally, a reconstruction algorithm is employed to combine the information from the conventional and TP-equipped temporal imaging systems into a higher resolution image pulse.\\
The rest of the paper is organized as follows: 
in section \ref{Theo}, a brief discussion on the low-pass behaviour of both spatial and temporal imaging systems has been provided. After that, paraxial (narrow-band) violation issue in imaging systems is explained in order to analyse the limitation of using time-lens array for resolution enhancement. After introducing the resolution issue in imaging systems, the proposed method to handle this problem is discussed in section 3. This method is verified through numerical simulations in section 4 and finally, the conclusion is provided in section 5.
\section{Theoretical background}\label{Theo}
\begin{figure}[htbp]
\centering
\includegraphics[width=\linewidth]{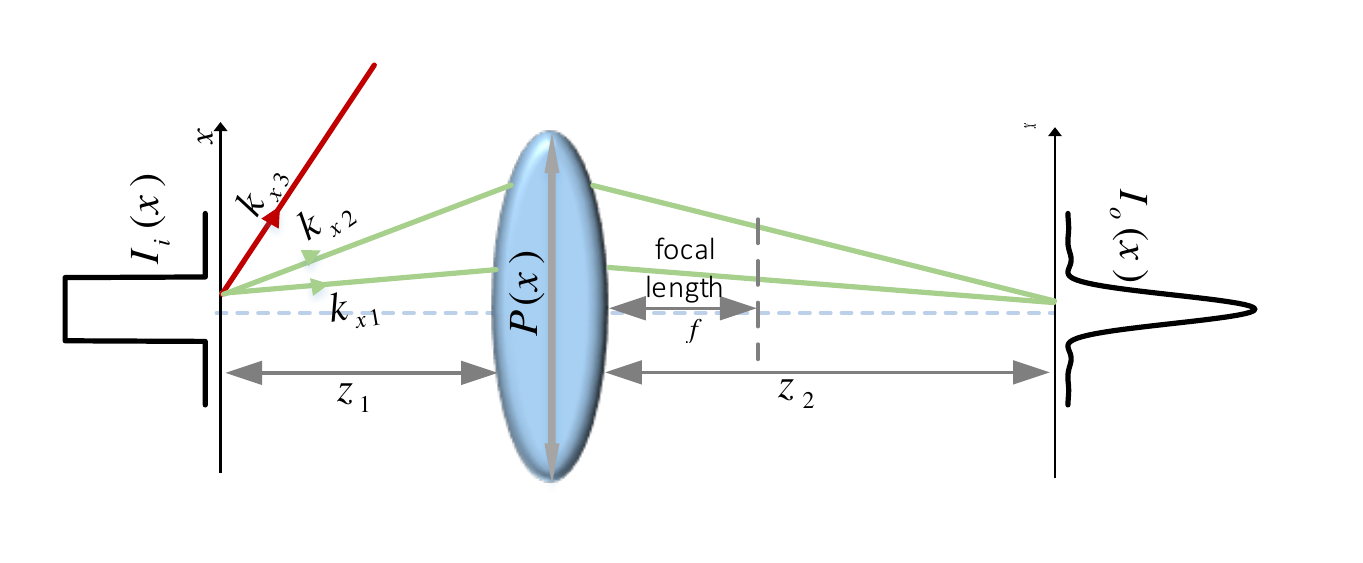}
\caption{Resolution limit due to the finite aperture of the lens. $k$ vectors with larger angles or higher spatial frequencies do not enter the lens. Missing high frequencies causes blurred image at the output.}
\label{fig1}
\end{figure}

\begin{figure}[ht!]
\centering
\includegraphics[width=\linewidth]{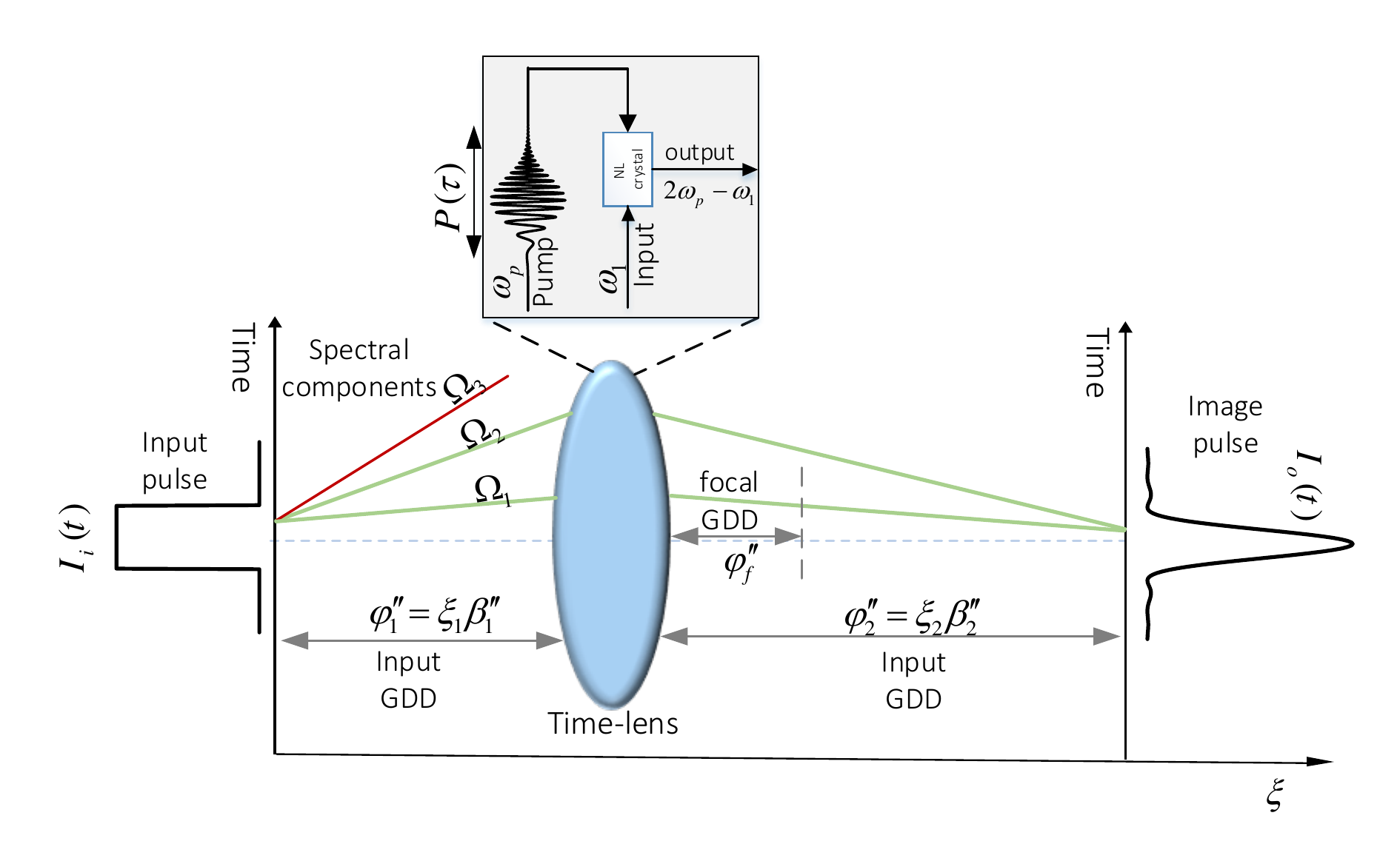}
\caption{In a temporal imaging system, temporal rays with higher spectral components, $\Omega$, experience more time delay than the lower frequencies. This leads to missing of higher frequencies at the output image because the time delay of these high-frequency components is more than the TL's aperture.}
\label{fig2}
\end{figure}

The most simple but intuitive imaging system both in spatial and temporal domains is composed of two diffractive (dispersive) mediums that are just before and after a lens (TL). Under image formation condition which is $1/{z_1}+1/{z_2}=1/f$ for spatial and $-1/{\phi_1''}+1/{\phi_2''}=1/{\phi_f''}$ for temporal imaging systems (with time-lenses based on four wave mixing (FWM) process), the image of the input appears on the image plane. Here $z_1$ is the distance between the lens and object and $z_2$ is the distance between the lens and the image in the spatial set up (Fig. \ref{fig1}). Furthermore, $\phi_1''$ and $\phi_2''$ are the group-delay-dispersions (GDD) of the first and second dispersive mediums in the temporal setup (Fig. \ref{fig2}). The variables $f$ and $\phi_f''$ represent the focal length and focal GDD of the spatial and temporal imaging systems, respectively. In an ideal imaging system, which is quite mathematical rather than physical, the exact replica of the input will appear on the image plane, however, the resolution limit is the most deterrent cause of nonideality. In the next two subsections, we are going to discuss resolution limit in both temporal and spatial cases.

\subsection{Spatial imaging system}\label{sis}
Fig. \ref{fig1} represent the resolution issue in a spatial imaging system. As can be noticed from this figure, the output of the imaging system for a 1-D rectangular input has become blurred. This blurring effect, as discussed mathematically in this section, is because of the finite aperture of the lens which makes the overall imaging system act as a low-pass filter. This means that the higher order angular spectral components do not experience the lensing effect (being imparted by a quadratic phase) and therefore they do not contribute to the image formation. In Fig. \ref{fig1}, $k_{x1}$ and $k_{x2}$ represent wave-vectors which have passed through the lens aperture. These frequency components have a lower value of projection along lateral, $x$, direction representing smaller frequency components. On the other side, wave vector $k_{x3}$, which has a larger projection along the lateral direction and thus brings the information of higher frequency components of the object has been blocked by the finite aperture of the lens. The systematic equation which describes this effect includes the pupil function
$P(x,y)$ as \cite{goodman}:
\begin{equation}\label{eq1}
\tilde{h}(u,\nu )=\int\limits_{-\infty }^{\infty }{\int\limits_{-\infty }^{\infty }{P(\lambda {{z}_{2}}\tilde{x},\lambda {{z}_{2}}\tilde{y})}}\exp [-j2\pi (u\tilde{x}+\nu \tilde{y})]d\tilde{x}d\tilde{y},
\end{equation}
where $\tilde{h}(u,\nu )$ is the point spread function (PSF) of the imaging system and $\tilde{x}=\frac{x}{{\lambda}{z_2}}$, $\tilde{y}=\frac{y}{{\lambda}{z_2}}$, $\tilde{h}={\frac{1}{M}}h$, and $M=\frac{z_2}{z_1}$. Assuming the pupil function to be a rectangular function of $x$ and $y$, the frequency response of the spatial imaging system, $\tilde{H}(k_u,k_v)$, could be obtained through Fourier transforming the PSF function represented in Eq. \ref{eq1}:
\begin{equation}\label{feq1}
\tilde{H}(k_x,k_y)=P({\lambda}{z_2}{k_x},{\lambda}{z_2}{k_y})
\end{equation}
This equation clearly represents the low-pass filtering effect caused by the finite aperture of the lens. This effect results in the high-frequency content of the object not to contribute to the formation of the image, which degrades the image resolution. As it will be discussed with more details in the next section, in order to enhance the resolution, the imaging structure and technique should be modified in a manner to pave the way for high-frequency components to reach the lens's aperture.\\

\subsection{Temporal imaging system}\label{tis}
In temporal imaging systems, the temporal aperture of the TL (both FWM based and EOM based TL) is limited in time. In an FWM based TL, the lens's aperture is limited to the temporal duration of the chirped pump pulse and therefore, the output of the TL can be written as
\begin{equation}\label{temp pupil}
  A_2(\tau)=A_1(\tau)H(\tau)P(\tau),
\end{equation}
where $H(\tau)$ is the pump's phase function, $A_1(\tau)$ is the input's pulse envelope, and $P(\tau)$ is the temporal aperture of the TL. Therefore, analogous to Eq. \ref{eq1}, the impulse response of the temporal imaging system at time $\tau$ to an impulse at time ${\tau}_0$ with a finite temporal aperture, $P(\tau)$, is $h(\tau)=c(\tau)\tilde{h}(\tau-M{\tau}_0)$, where \cite{Bennette20000}
\begin{equation}\label{ctau}
  c(\tau)=1/{\sqrt{M}}\exp(\frac{-i{\tau}^2}{2M{\phi_f}''}),
\end{equation}
and
\begin{equation}\label{impulse of temporal}
  \tilde{h}(\tau-{M{\tau_0}})=\frac{1}{2\pi {{\phi}_{2}}''}\int\limits_{-\infty }^{+\infty }{P({\tau }')\exp [\frac{-i{\tau }'(\tau -{{{{\tau }'}}_{0}})}{{{\phi}_{2}}''}]}d{\tau }'.
\end{equation}
Taking the Fourier transform of the above equation, the frequency response of the temporal imaging system, $H(\omega)$, can be written as
\begin{equation}\label{freqH}
  \tilde{H}(\omega)=CP({{\phi_2}''}\omega)
\end{equation}
where $C$ is a phase term resulting from the multiplication of the Fourier transform of Eq. \ref{ctau} and the phase term created by the temporal shift in the Fourier transform of Eq. \ref{impulse of temporal} , ${\phi_2}''$ is the GDD of the second dispersive medium, and ${\phi_f}''$ is the GDD of TL.
Comparing Eq. \ref{feq1} and Eq. \ref{freqH}, one can observe that if the pupil function has a finite size or finite duration, as is always in a real structure, then the imaging system, whether spatial or temporal, will behave as a low-pass filter.\\
In order to further investigate the resolution limit of a temporal imaging system, we explain the notion of the time ray which was first introduced by Bennett and Kolner in the extensive works that they have done in this area \cite{Kolner2001}. In time-ray visualization, every frequency component of a pulse has a distinct space-time path. This illustration which is shown in Fig. \ref{fig2} is based on the fact that different frequencies in a dispersive media propagate with different velocities. Higher frequencies have time delay bigger than the lower frequencies in a particular length because lower frequencies travel faster than the higher ones in a normal dispersive medium. Therefore, different frequencies have different slopes in this space-time representation. This also can be described mathematically by first expanding the group delay as \cite{Kolner2001}
\begin{equation}\label{m1}
\begin{split}
t_g(\omega) & =\frac{z}{v_g(\omega)}=z\frac{d\beta(\omega)}{d\omega} \\
     & = z(\beta'+(\omega-\omega_0)\beta''+(\omega-\omega_0)^2\frac{\beta'''}{2!}+\cdots).
\end{split}
\end{equation}
For a particular base-band spectral component $\Omega=\omega-\omega_0$, the group delay with respect to traveling-wave coordinates is
\begin{equation}\label{m2}
  \begin{split}
     \tau_g(\Omega) & = t_g(\omega)-t_g(\omega_0) \\
       & = \Omega{\xi}{\beta{''}}+{\Omega^2}\frac{\xi{\beta{'''}}}{2!}+{\Omega^3}\frac{\xi{\beta{''''}}}{3!}+\cdots \\
       & = \Omega{\phi{''}}+ {\Omega^2}\frac{{\phi{'''}}}{2!}+{\Omega^3}\frac{{\phi{''''}}}{3!}+\cdots.
  \end{split}
\end{equation}
\begin{figure}[!hb]
\includegraphics[width=\linewidth]{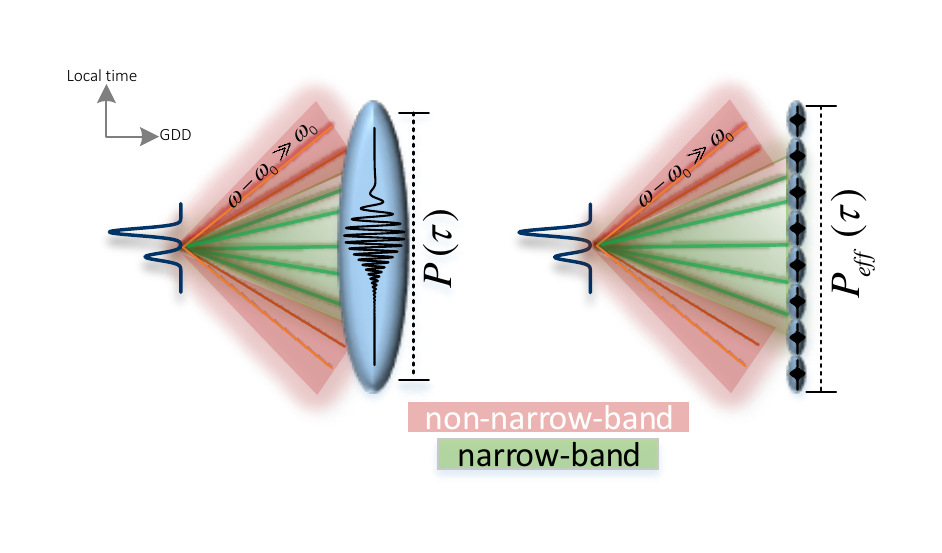}
\caption{Time ray diagram showing the narrow-band limitation of systems with large aperture time lens (left) and time-lens array (right). High spectral components violate the narrow-band condition ($({\omega-\omega_0})/{\omega_0}\gg{1}$) which is an assumption for image formation in temporal systems.}
\label{figparaxial}
\end{figure}
 When a frequency component enters a dispersive media in an initial time $\tau_{in}$, then it exits that media at $\tau_{out}(\Omega)=\tau_{in}+\tau_g(\Omega)$. Therefore, we can define the slope of time-rays as $d\tau(\Omega)/d{\phi}''=d\tau_g(\Omega)/d{\phi}''$ which can be written as
 \begin{equation}\label{m3}
 \frac{d{\tau}(\Omega)}{d\phi{''}}=\Omega{[1+\frac{\Omega}{2!}\frac{\beta'''}{\beta''}+\frac{\Omega^2}{3!}\frac{\beta''''}{\beta''}+\cdots+\frac{\Omega^{n-2}}{(n-1)!}\frac{\beta^{(n)}}{\beta''}]}.
 \end{equation}

\begin{figure*}[!ht]
  \centering
  \includegraphics[width=\linewidth]{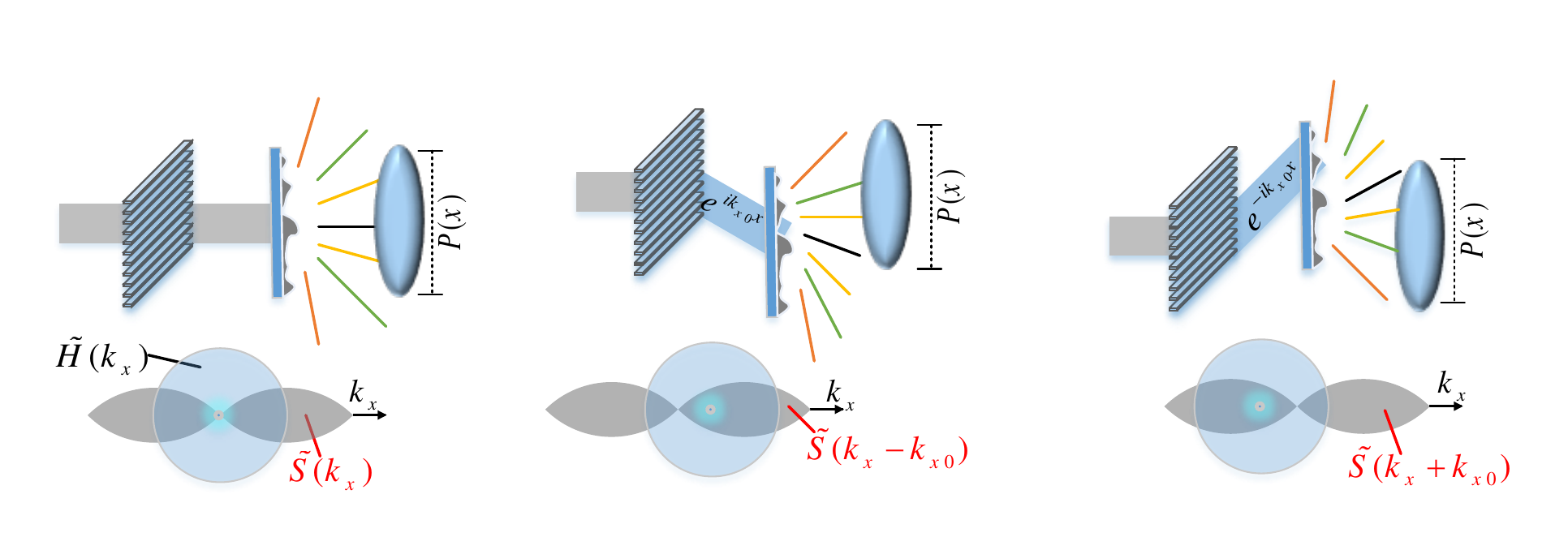}
  \caption{Structured illumination based super-resolution in a spatial imaging system. The grating produces oblique incidences on the object in order to shift the angular spectrum of the object. Therefore, those areas of the image's spectrum which were band-limited by the system, now can contribute to the output image. $\tilde{S}(k_x)$, $\tilde{S}(k_x+k_{x0})$, and $\tilde{S}(k_x-k_{x0})$ represent the original, right shifted, and left shifted spatial angular spectrum of the input, respectively. $\tilde{H}(k_x)$ is the frequency response of the imaging system and $P(x)$ is the pupil function of the lens.}\label{prop s}
\end{figure*}

Because in this article we are not interested in the effects of aberrations produced by a dispersive medium (which can be described by taking into account the higher terms of Taylor expansion of $\beta(\omega)$ where $\beta(\omega)$ is the frequency dependent wave-number), only first term of Eq. \ref{m3} is kept and others are left out. Therefore, when a baseband spectral component, $\Omega$, enters a dispersive medium and propagates a distance $\xi_0$, it will have a GDD equal to $\phi_0''=\xi_0\beta''$ . Next, having the slope of its time-ray diagram using Eq. \ref{m3}, we are able to calculate the time delay of that frequency component.\\
Using this concept, the resolution limit of the TL is quite easy to understand because higher frequencies do not have enough time to reach to the finite temporal aperture of the TL. In other words, the frequency components that have a group delay bigger than the aperture of the TL cannot pass through the lens and lensing effect (imparting quadratic temporal phase) does not apply to them. In the next section, we have used this concept to understand the proposed method more visually.\\
One tempting idea in order to have better resolution is to increase the band-width of the imaging system using a lens with a wider aperture. Nonetheless, it suffers from some crucial problems among them the most important one is paraxial approximation (in spatial optics). It should be noted that the whole idea of imaging is based on the paraxial approximation which is described by geometric optics. In geometrical optics, the paraxial approximation is a small-angle approximation used in Gaussian optics and ray tracing of light through an optical system. As shown in Fig. \ref{figparaxial}, paraxial ray is a ray which makes a small angle to the optical axis of the system, and lies close to the axis throughout the system. Paraxial approximation is a first order approximation. To deal with aberrations, higher order approximations or non-paraxial geometrical optics are needed. The same issue occurs when using a lens array instead of a single lens in imaging systems. That is because in a lens array, the constituent lenses collectively act like a big aperture lens. The above discussion also holds in temporal imaging systems. In that case, however, image formation occur under narrow-band assumption for the input pulse. In fact, increasing the resolution of a temporal imaging system using a large temporal aperture time-lens or an array of time-lenses, which as mentioned earlier was proposed by \cite{zhan}, would lead narrow-band time rays to become captured by the time-lens or time-lens array.\\
The proposed method in this paper exploits time prisms to shift higher spectral components before passing through each time-lens in order to be consistent with the narrow-band condition. \\
\section{Proposed method}\label{proposed method}
Instead of using a large aperture time lens or a time lens array, our proposed method utilized a pair of time prisms in order to capture high frequency components of the input pulse in order to increase the bandwidth of the imaging system and therefore its resolution. Time prisms which are temporal counterparts of spatial prisms are more discussed in the subsection \ref{limitsub}. This technique is the dual of the idea used in structured illumination microscopy as a successful super-resolution method in spatial domain which is by the first time proposed in this paper in the time-domain.\\
\begin{figure}[!hb]
\centering
  \includegraphics[width=3.5in]{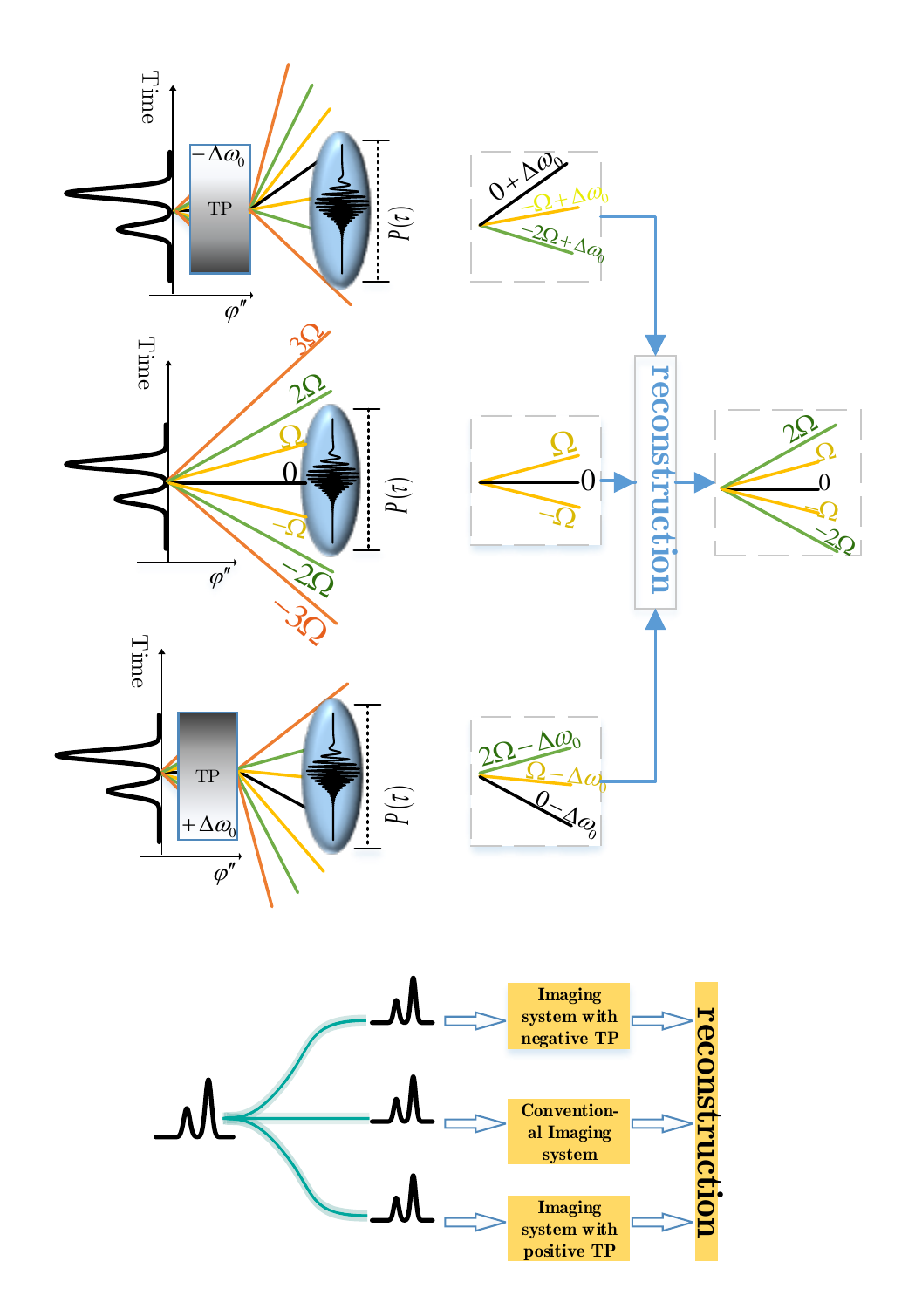}
 \caption{Top: schematic of the proposed method using time ray visualization. Spectral components $0$, $\pm{\Omega}$, $\pm{2\Omega}$, and $\pm{3\Omega}$ are represented as sample spectral components. Because of dispersion, these components have different velocities and therefore, different time-delays. The amount of time-delay for higher frequencies (for example $3\Omega$) is more than the duration of TL's aperture. Hence, Similar to the spatial imaging systems, some of these rays do not pass through the finite aperture TL. Temporal prisms, therefore, are recruited to shift the spectra of the pulse. These TPs change the slope of time rays in a way that fit the ray into the TL's aperture. After applying the first TP, ($+\triangle{\omega}_0$), the time ray $2\Omega$ can pass through the TL and after applying the second TP ($-\triangle{\omega}_0$), the $-2\Omega$ time ray can be passed through the TL. $P(\tau)$ is the temporal pupil function of the TL. Bottom: by making 3 or more copies of the input pulse through and optical coupler, the whole imaging system can operate in a single shot manner.}
  \label{fig4}
\end{figure}
\begin{figure}[!hb]
\includegraphics[width=7cm]{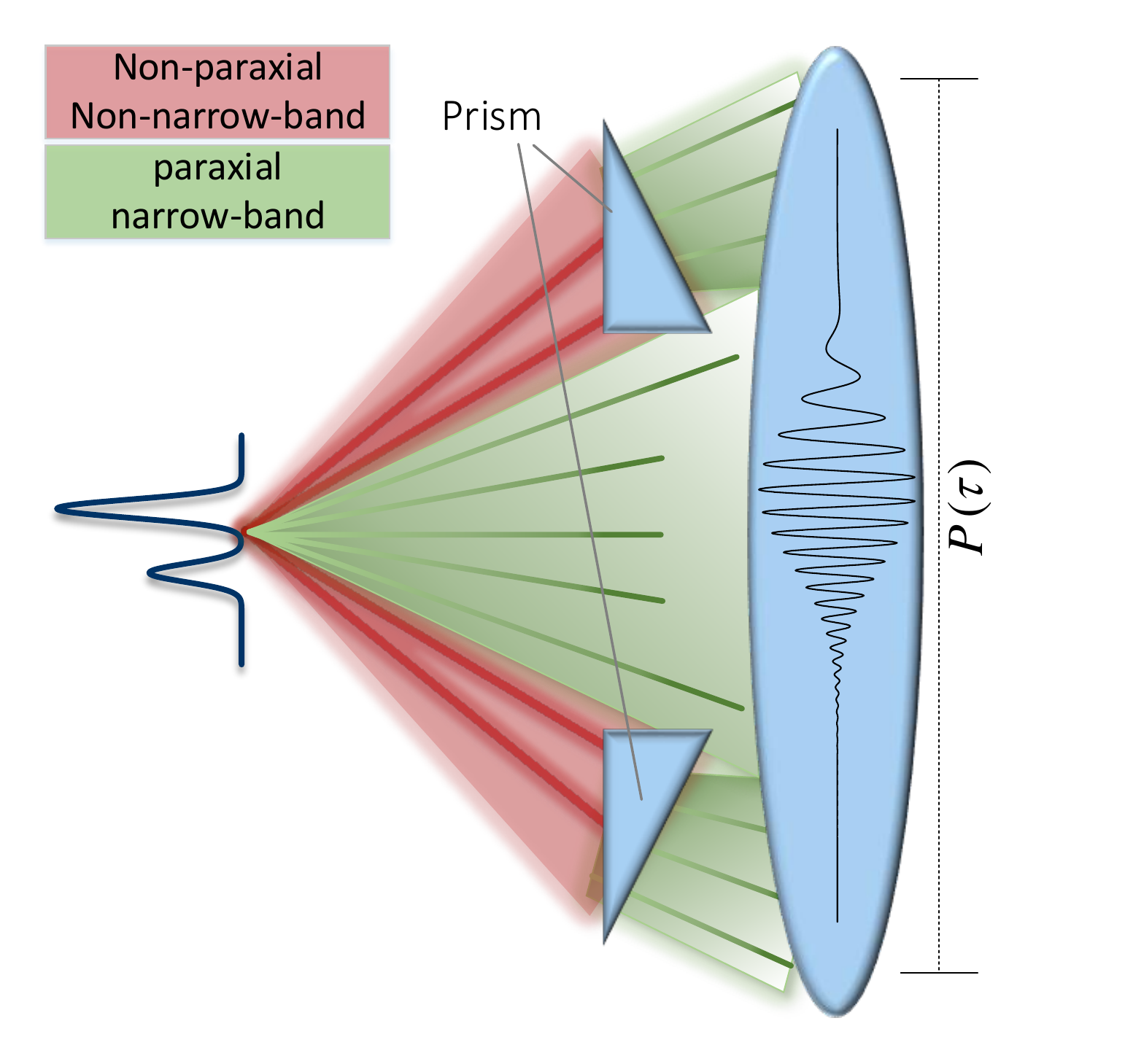}
\caption{a}
\label{proposed22}
\end{figure}
In order to discuss the proposed method, some basic ideas behind spatial SISR should first be noticed to establish the analogy. In every spatial imaging system, the detected intensity of the image can be written as \cite{Suxena}
\begin{equation}\label{prop1}
  D(x)=I_s(x){\ast}h(x),
\end{equation}
where the ${\ast}$ denotes the convolution operator and $I_s(x)$, $h(x)$, and $D(x)$ represent the light intensity of the object, PSF, and intensity of the image, respectively. Here, for simplicity, we have assumed that the object and the imaging system is one dimensional. The light intensity emitted from the sample $I_s(x)$ depends on the incoming light intensity ($I_{in}$) and the object's structure ($S(x)$) and can be written as

\begin{equation}\label{prop2}
  I_s(x)=I_{in}S(x).
\end{equation}
Transforming Eq. \ref{prop1} into frequency domain and substituting $I_s(x)$ from Eq. \ref{prop2}, the frequency spectra of the output's light intensity will be
\begin{equation}\label{prop3}
  \tilde{D}(k_x)=[\tilde{I}_{in}(k_x){\ast}\tilde{S}(k_x)]\tilde{h}(k_x),
\end{equation}
where $\tilde{h}(k_x)$, $\tilde{S}(K_x)$, and $\tilde{I}_{in}(k_x)$ are the frequency representations of the PSF, Object's structure, and the incident's light intensity, respectively. In the case of a uniform plane wave illumination ($\tilde{I}_{in}(k_x)={I_0}{\delta}(k_x)$) on the object, the spectrum of image becomes
\begin{equation}\label{prop4}
  \tilde{D}(k_x)=I_{0}\tilde{S}(k_x)\tilde{h}(k_x).
\end{equation}
The important point to notice here is that because $\tilde{h}(k_x)$ is limited in frequency domain, the imaging system, as discussed in section \ref{sis}, would be a low-pass filter. On the other hand, if we use a structured illumination rather than a uniform plane wave, for example a sinusoidal illumination such as
\begin{equation}\label{prop5}
  I_{in}={I_0}[1+m/2(e^{j({k_{x0}}x+\phi)}+e^{{-j(k_{x0}x+\phi)}})],
\end{equation}
the frequency spectra of the output's light intensity will become
\begin{equation}\label{prop6}
\begin{split}
   \tilde{D}(k_x) & =I_0\tilde{h}(k_x)\{\tilde{S}(k_x)+({m/2}){\exp(j\phi)}\tilde{S}(k_x-k_{x0}) \\
     & +({m/2}){\exp(-j\phi)}\tilde{S}(k_x+k_{x0})\}.
\end{split}
\end{equation}

\begin{figure}[ht]
  \centering
  \includegraphics[width=3in]{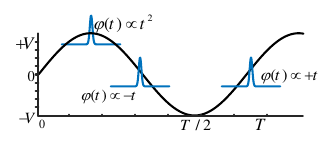}
\caption{Different functionalities of an electro-optical modulator driven with a sinusoidal voltage. Positive linear, negative linear and negative quadratic phase shifts are shown in this figure.}
  \label{eom}
\end{figure}

According to the second and third terms in the above equation, higher frequencies of the object's spectrum can be fell down in the pass-band of the imaging system provided that the incident light is structured as Eq. \ref{prop5}. In Eq. \ref{prop6}, the output's spectrum has three terms. These terms correspond to the detected spectrum of the object with $0$, $+k_{x0}$, and $-k_{x0}$ spectral shifts, passed through the low-pass frequency response of the system. In other words, the first term contains the lower frequency part of the object's reflectivity pattern while the second and third terms include the high-frequency components of the object's reflectivity pattern shifted and passed through the imaging system's frequency response. After detecting these three components of the input's spectrum, a computational reconstruction needs to be done in order to attach these three parts of the object's spectrum and reconstruct the wide-band spectrum of the object. This technique is illustrated schematically in Fig. \ref{prop s}. In this figure, a grating device is exploited to make oblique illuminations on the object (three oblique illuminations has been depicted in this figure to be comparable with equation \ref{prop5}). The first illumination is normal to the object resulting in the passage of lower frequency components, which have $k$-vectors with less inclination, through the lens. The second and third illuminations are inclined with an angle $\theta$ resulting in rotation of more oblique $k$-vectors which correspond to higher frequency terms, to be inclined and thus pass through the lens's aperture. Therefore, the second and third illuminations result in imaging the more inclined $k$-vectors and thus higher frequency components of the object. One important point to be noticed is that either object or illuminations in the above theoretical analysis have been assumed to be one dimensional. However, in a real imaging system, the object is two-dimensional with transparency $S(x,y)$ and the oblique illuminations are plane-waves varying in both $x$ and $y$ directions, making spectral shifts of value $\pm{k_{x0}}$ and $\pm{k_{y0}}$ to the angular spectrum of the object $\tilde{S}(k_x,k_y)$.\\
This technique has an analogue in electronics known as modulated wide-band convertor (MWC), which is designed for sampling a wide-band signal with sub-Nyquist rate \cite{mishali}. By modulating the input with several periodic signals in and then passing each branch through a low-pass filter, wide-band spectrum of the original signal can be achieved with a combination of the outputs of these branches. SISR can be viewed as an optical counterpart of MWC in which a grating is responsible for modulating the object's light intensity by periodic light and the finite aperture of the lens acts as a low-pass filter.\\
Taking advantage from the discussed method in spatial imaging, we propose a similar trick in temporal imaging systems to pass the high-frequency components of the input pulse to the output even though the system is a low-pass filter. In order to do so, the spectra of the input pulse must be shifted analogous to the spatial case discussed above. To shift the input's spectra, a temporal prism has been exploited. Just likewise usual spatial prisms that shift the $k$ vector of the light-wave, temporal prisms are used to shift the frequency components $\Omega$ of the input temporal pulse. The role of the time-prisms in this proposed method is similar to the role of the gratings in spatial SISR. Illustration of this method is shown in Fig. \ref{fig4}. Comparing the spatial and temporal super-resolution techniques in this figure and Fig. \ref{prop s}, one point needs to be noticed. In the proposed temporal super-resolution procedure, time-prisms are right after the input while in spatial SISR, the grating is behind the object, making a structured illumination on the object. It can be discerned from the Eq. \ref{prop6} that the periodic illumination leads to the shifting of the object's spectrum. Therefore, because there is no illumination in the temporal imaging system as in spatial ones, the time prisms must shift the spectrum directly. These two settings (shifting the spectrum by time prisms or by structured illumination) do not differ mathematically as both shift the spectrum of the input.\\
In Fig. \ref{fig4} the concept of time-rays has been employed to visualize the proposed method. As can be seen in this figure, two time prisms have been exploited to rotate the temporal rays in two sides resulting in a frequency shift of the input in both sides. The $+\triangle{\omega_0}$ TP causes the $2\Omega$ time-ray to be captured by the TL, while the $-\triangle{\omega_0}$ TP leads to the capturing of $-2\Omega$ time-ray by the TL. Therefore, these two time-rays which were filtered in the conventional temporal imaging system (without time prisms) due to the finite pupil function of the TL, now can contribute to the imaged pulse at the output.\\
\subsection{The limit in resolution improvement}\label{limitsub}
Before discussing the limitation in the resolution enhancement, we first discuss the structure of time prisms. Time-prisms, which are usually made of electro-optic modulators (EOM), are the temporal counterparts of conventional spatial prisms. The linear phase term imparted by the time prism to the input pulse is realizable by an EOM driven with a sinusoidal voltage.\\
In an EOM the change in refractive index due to the voltage can be written as $\triangle{\phi_{EOM}}=\triangle{n}{\frac{2\pi}{\lambda}}L=-\frac{L{{n}_3^3}r_{33}\Gamma}{\lambda{\omega}\pi{V}}=\frac{\pi{V}}{V_{\pi}}$, where $\Gamma$ accounts for the inhomogeneity of the modulator, $r_{33}$ is the coupling constant between electric field and refractive index, $n_3$ is refractive index with no electric field applied, and $V_{\pi}$ is known as $\pi$ or half-wave voltage \cite{LJW}. By applying a sinusoidal drive voltage $V=V_Dsin(\omega_Dt)$ and writing for the linear portion of the sinusoidal function, the induced phase is
\begin{equation}\label{inducedphase}
  \triangle{\phi{(t)}}=\frac{\pi{V_D}}{V_\pi}{\sin}({\omega}_D{t})\simeq{\triangle\omega_0}t,
\end{equation}
where $\omega_D$  and $V_D$ are the frequency and amplitude of the driving voltage. Obviously, since the refractive index of the EOM material is now linearly proportional to time $t$, the induced phase will also have a linear dependence on $t$ which is a necessity for implementing a time-prism. However, this phase shift is finite. Therefore, using the proposed method with a pair of time prism would have a limited resolution enhancement because of the limited shift done by the time prism to the input's spectra and therefore, limited increase in the effective bandwidth of the imaging system. However, one might suggest utilizing more than two time prisms in order to capture more frequency content of the input pulse. This might seem a plausible idea, however, as shown in Fig. \ref{limit}, although multiple time prisms would enable those high frequency rays to enter the time-lens, they would no more meet the narrow-band condition. That means, since the higher frequencies are already out of narrow-band condition and the phase shift imparted by the time prisms are finite, they still remain in the non-narrow-band regime, which do not contribute in the image formation.  

\section{Simulations and results}\label{sim}
\begin{figure*}[ht]
  \centering
  \includegraphics[width=14cm]{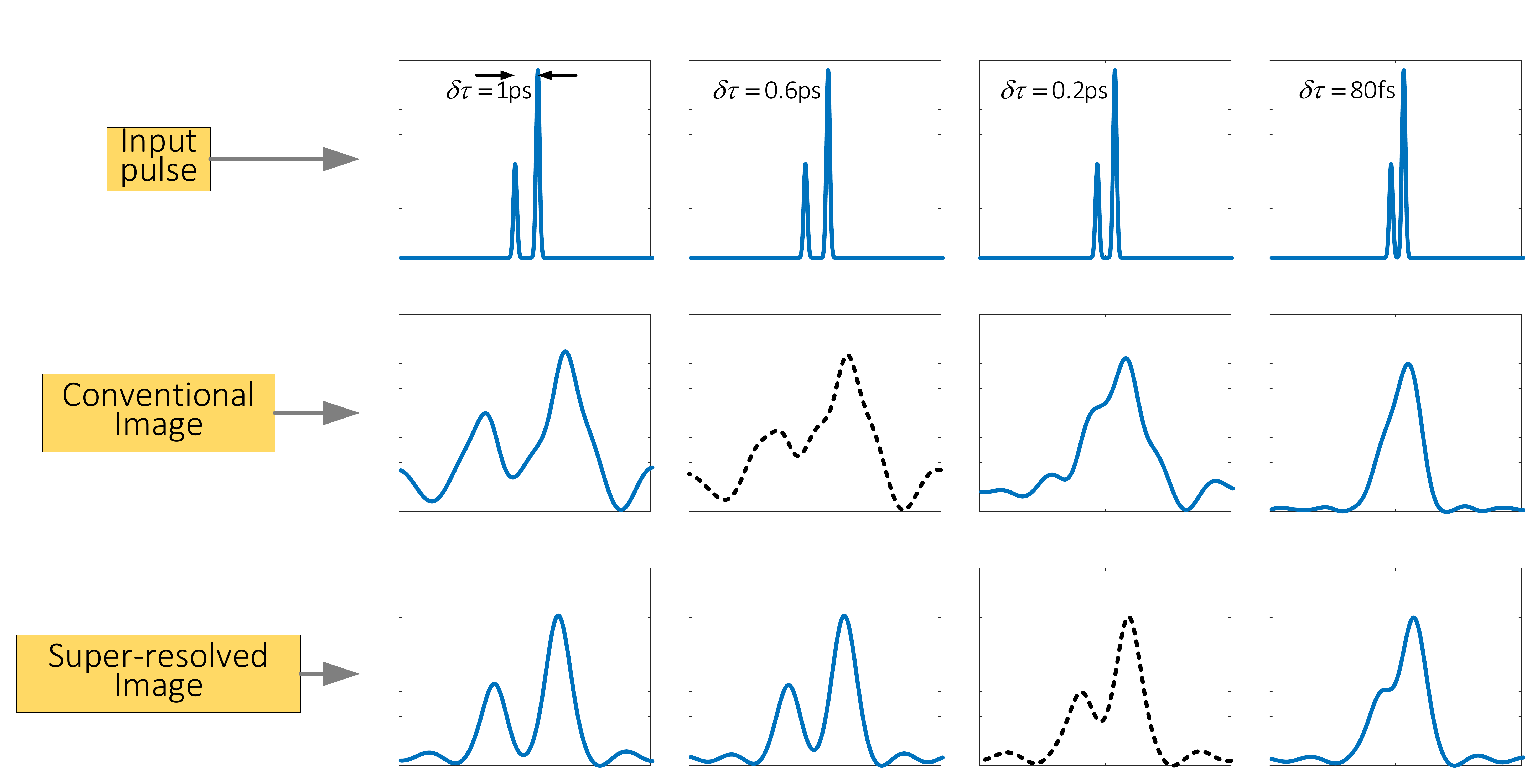}
\caption{Visualisation of resolution enhancement for different pulse separations.}
  \label{7}
\end{figure*}
\begin{figure}
  \centering
  \includegraphics[width=4cm]{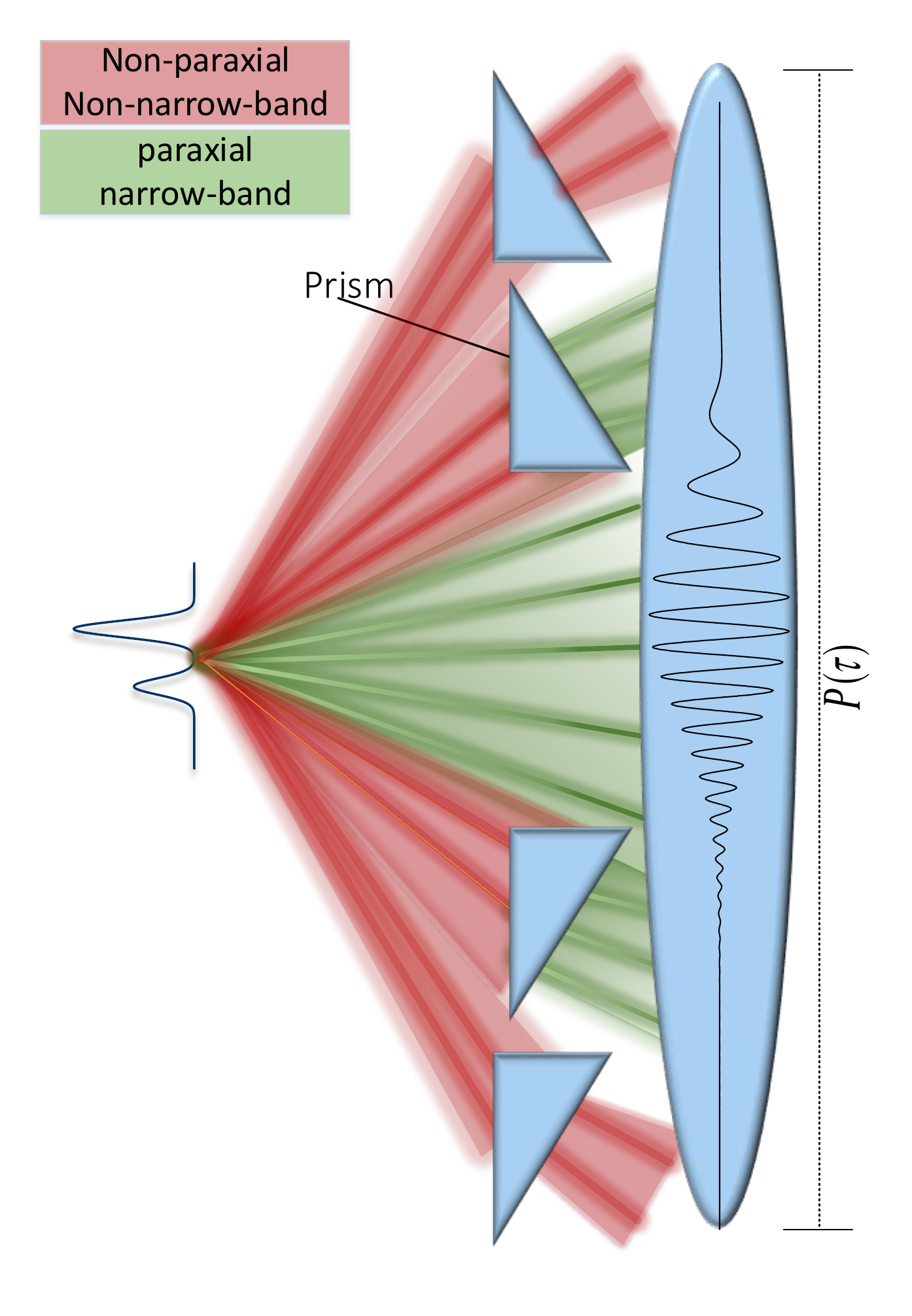}
\caption{Limitation.}
  \label{limit}
\end{figure}
In this section, numerical verification of the proposed method has been discussed. The temporal imaging system in this simulation is assumed to be an FWM-based TL with a Gaussian aperture (the temporal envelope of the pump pulse of the FWM process is Gaussian) which has a temporal duration of $P(\tau)=6 ps$. The magnification factor of the considered temporal imaging system is $M=10$. Besides, the GDD of the second dispersive medium is $\phi_2''=10ps^2$. Therefore, according to this magnification factor and also the image formation condition for a FWM-based time lens (-1/$\phi_1''$+1/$\phi_2''$=1/$\phi_f''$ \cite{Bennette2000}), we have $\phi_1''=1ps^2$ and $\phi_f''=-1.1ps^2$. The bandwidth of the imaging system (according to Eq. and the choice for $\phi_2''$) is about $0.6 THz$. The minimum temporal width of the input pulse (shown in Fig. 6(a)) is $0.1ps$. The spectral width corresponding to this input pulse is about $7 THz$ \cite{saleh}, which is more than the imaging system's bandwidth. Therefore, higher frequency components would become filtered. Ideally, in order to image and resolve this pulse, we need an imaging system that can resolve the fastest temporal feature of this pulse. For better exploring this matter, an input signal consisting three pulses each with $0.1 ps$ temporal width and $0.2 ps$ and $0.4 ps$ separations is considered, as shown in Fig. 6(a). In Fig. 6(b), the blue curve is the image of the input signal with the considered conventional temporal imaging system, which, as shown, could not resolve non of the $0.2 ps$ and $0.4 ps$ separations. Based on the numerical simulation and also theoretical analysis, these values of resolvability requires and imaging system bandwidth more than $1.5THz$, which is larger than the frequency bandwidth of the considered conventional system, i.e., $0.6 THz$. Now, by introducing two time prisms with positive and negative phase shifts equal to $500 GHz$, the output image would be as the red curve in Fig. 6(b). This amount of phase shift imparted by the time prisms are nowadays experimental tested and available \cite{500GIG1,500GIG2}. Introduction of time prisms to the system is analogous to have a new bandwidth for the whole temporal imaging system which is wider than before. Therefore, considering $|\Delta_{\omega{0}}|=0.5 THz$ frequency shifts both for negative and positive time prisms, the equivalent bandwidth of the system will become $0.6+2(|\Delta_{\omega{0}}|)=1.6 THz$, which is adequate for resolving both $0.2 ps$ and $0.4 ps$ pulse separations, as will be discussed in the following.\\
\begin{figure}[ht]
  \centering
  \includegraphics[width=\linewidth]{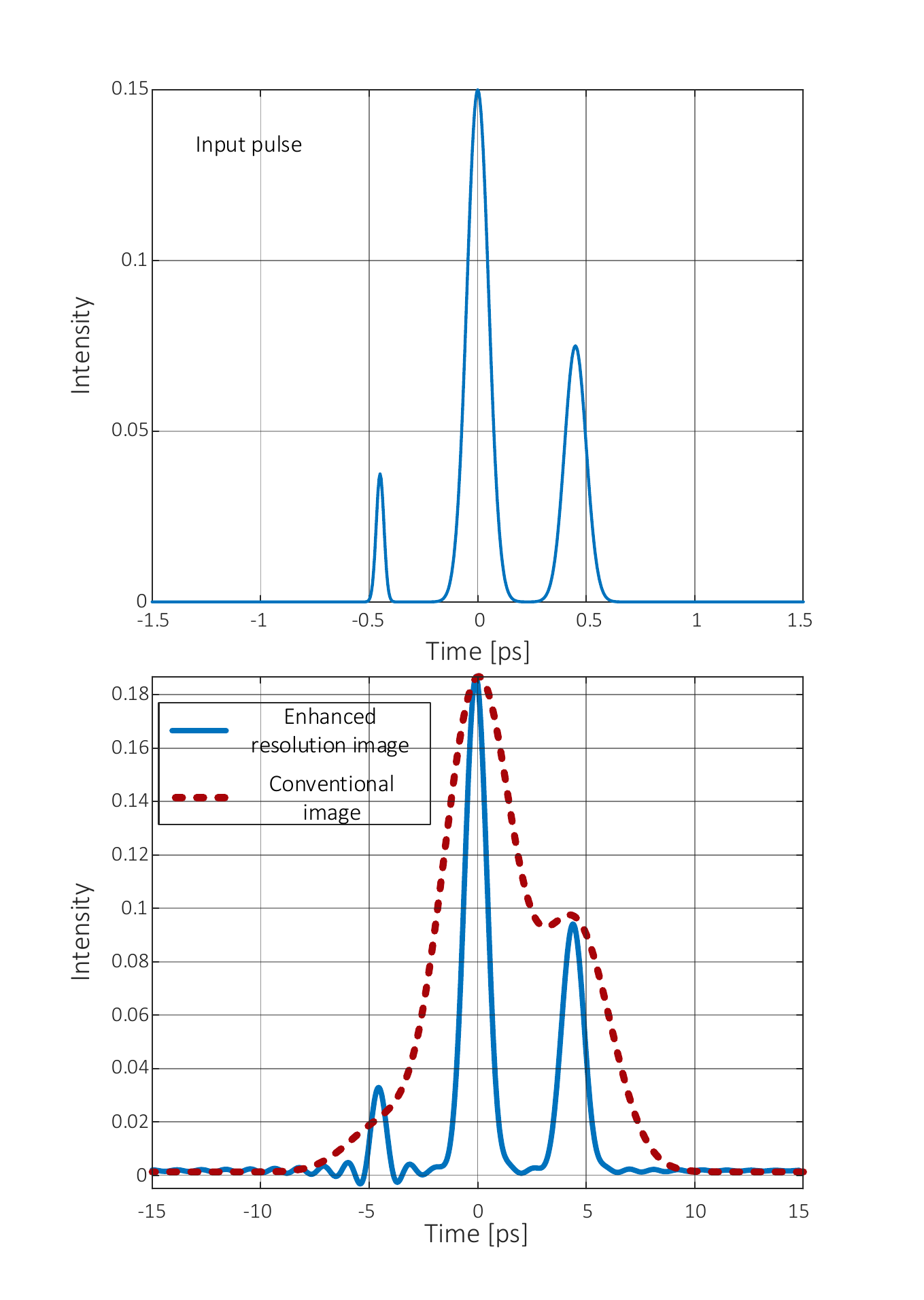}
\caption{Simulation results for the proposed method for a pulse consisted of three peaks.}
  \label{6}
\end{figure}
\begin{figure}[!hb]
  \centering
  \includegraphics[width=\linewidth]{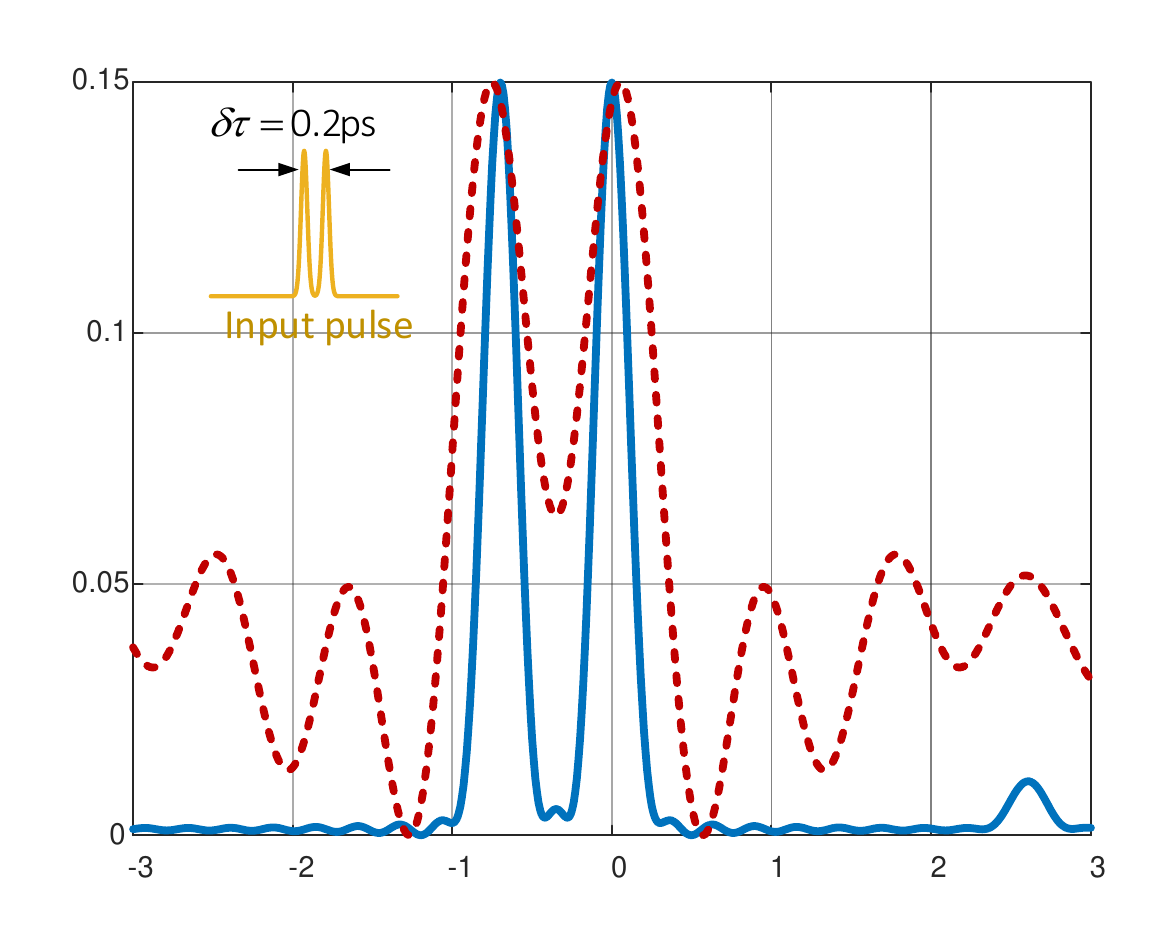}
\caption{Effect of increasing number of time prisms.}
  \label{8}
\end{figure}
The amount of increased resolution has been verified as in Fig. 7. In this figure, two pulses, each with $0.1 ps$ temporal widths are getting closer to each other in time until in their image, the peaks cannot be distinguished. According to these simulations, in the conventional temporal imaging system (without time prisms), the minimum pulse separation between two pulses that could be resolved by the system is $0.5 ps$. On the other hand, in the case where the proposed method is used, this minimum separation decreases to $0.2 ps$. Thus, according to this criterion for defining temporal resolution, it is clear to see that the resolution can also be explained by how much the bandwidth of the time prism-equipped system is wider than the bandwidth of the conventional system. This ratio, which is 2.66 (the ratio of $1.6 THz$ to $0.6 THz$), is almost consistent with the resolution improvement factor obtained from the simulations.\\
Likewise the final step in spatial SISR, a reconstruction procedure must be applied to retrieve the information from all these three (or more) branches (Fig. \ref{fig4}). Firstly, we write three obtained spectrums as \cite{Kaveh}
\begin{eqnarray}\label{recon1}
  S_1(\omega) &=& F(\omega).P(\omega) \\
  S_2(\omega) &=& F(\omega-\triangle{\omega_0}).P(\omega)  \\
  S_2(\omega) &=& F(\omega+\triangle{\omega_0}).P(\omega)
\end{eqnarray}
where $P(\omega)$ represents the frequency response of the temporal imaging system which is related directly to the lens's aperture as in Eq. \ref{freqH}, and $F(\omega)$ is the input's spectrum. Here, it has been assumed that $P(\omega)$ is the full-width at half maximum (FWHM) spectral window of the Gaussian aperture. $S_2(\omega)$ and $S_3(\omega)$ contain the high-frequency contents of the spectrum and $S_1(\omega)$ accommodate the low-frequency information of the input's spectra. Given that $S_2(\omega)$ and $S_3(\omega)$ are the shifted high-frequency components to the center, they should be shifted to their right location in the spectrum as follows:

\begin{eqnarray}\label{recon2}
  {\tilde{S}}_2(\omega) &=& S_2(\omega+\triangle{\omega_0}) \\
   {\tilde{S}}_3(\omega) &=& S_3(\omega-\triangle{\omega_0})
\end{eqnarray}
Although ${\tilde{S}}_2(\omega)$ and ${\tilde{S}}_3(\omega)$ contain the high-frequency information, they have some overlaps with $S_1(\omega)$. Therefore, if we simply add ${\tilde{S}}_2(\omega)$, ${\tilde{S}}_3(\omega)$, and $S_1(\omega)$ together, the overlapping areas will be counted twice which leads to a distorted image spectrum. In order to cut the overlapping areas, these three spectrums should be filtered as follows:
\begin{eqnarray}\label{recon3}
  {\tilde{S}_2}^{HP}(\omega) &=& {\tilde{S}}_2(\omega).(1-P(\omega)) \\
  {\tilde{S}_3}^{HP}(\omega) &=& {\tilde{S}}_3(\omega).(1-P(\omega))
\end{eqnarray}
where $1-P(\omega)$ is a high-pass filter, representing the frequencies out of the pass-band of the conventional temporal imaging system. Finally, we can reconstruct the wide-band image pulse by taking an inverse Fourier transform as:
\begin{equation}\label{last}
  I_0(t)={F}^{-1}{\{S_1(\omega)+{\tilde{S}}_2^{HP}(\omega)+{\tilde{S}}_3^{HP}(\omega)\}}
\end{equation}
The resultant temporal signal contains missed high-frequency components in the conventional system and thus is certainly of higher frequency bandwidth compared to the imaged pulse from conventional temporal imaging system. Fig. \ref{result} has been generated using this algorithm.

 After applying the discussed reconstruction algorithm to the captured data from the proposed super-resolved temporal imaging structure, the resolved output pulse was obtained (red curve in Fig. \ref{result}). Comparing with the original input pulse (yellow curve), the resolvability of the proposed structure becomes evident. The widening of the pulses is due to the lower resolution of the proposed structure than the pulses' width. In this Fig. \ref{result}, the blue curve is the unresolved image pulse from the conventional temporal imaging system which has been depicted in order to demonstrate the resolution enhancement of the proposed method.

\section{Conclusion}
A method similar to the conventional structured illumination super-resolution has been presented in order to increase the temporal resolution of a temporal imaging system. The proposed method utilizes a pair of time prisms to shift the spectral components of a temporal pulse as in the spatial structured illumination super-resolution technique, where oblique illuminations cause spectral shifting. Shifting the input's spectrum by time prisms causes to the contribution of the input's high frequency components to the formation of image and therefore, increasing the temporal resolution. Numerical proofs with practical parameters have been provided to test the proposed method. The suggested time prisms shift the spectrum by $\pm{2.3}$ THz and are made of electro-optic modulators with a sinusoidal driving voltage. The resolution limit of the assumed temporal imaging system has been enhanced by a factor of $3$ and the equivalent imaging bandwidth of the TP-equipped imaging system has been increased. After this study, an experimental realization of the proposed concept is possible.


\begin{thebibliography}{17}

\bibitem{kolner94}
Brian H. Kolner and Moshe Nazarathy, "Temporal imaging with a time lens," Optics Letters, vol. 14, no. 12, p. 630, Jun. 1989.


\bibitem{salem}
Reza Salem, Mark A. Foster, and Alexander L. Gaeta, "Application of space–time duality to ultrahigh-speed optical signal processing," Advances in Optics and Photonics, vol. 5, no. 3, p. 274, Sep. 2013.

\bibitem{howe}
J. van Howe and C. Xu, "Ultrafast optical signal processing based upon space-time dualities," Journal of Lightwave Technology, vol. 24, no. 7, pp. 2649–2662, Jul. 2006.


\bibitem{moticloak}
Moti Fridman, Alessandro Farsi, Yoshitomo Okawachi, and Alexander L. Gaeta, "Demonstration of temporal cloaking," Nature, vol. 481, no. 7379, pp. 62–65, Jan. 2012.


\bibitem{talbotazana}
José Azaña and M. A. Muriel, "Temporal Talbot effect in fiber gratings and its applications," Applied Optics, vol. 38, no. 32, p. 6700, Nov. 1999.


\bibitem{goodman}
J. W. Goodman, Introduction to Fourier Optics. McGraw-Hill, 1996.


\bibitem{L66}
W. Lukosz, "Optical Systems with Resolving Powers Exceeding the Classical Limit*," Journal of the Optical Society of America, vol. 56, no. 11, p. 1463, Nov. 1966.


\bibitem{Wilde16}
Jeffrey P. Wilde, Joseph W. Goodman, Yonina C. Eldar, and Yuzuru Takashima, "Coherent superresolution imaging via grating-based illumination," Applied Optics, vol. 56, no. 1, p. A79, Jan. 2017.


\bibitem{Flu}
J Vangindertael, R Camacho, W Sempels, H Mizuno, P Dedecker, and K P F Janssen, "An introduction to optical super-resolution microscopy for the adventurous biologist," Methods and Applications in Fluorescence, vol. 6, no. 2, p. 022003, Mar. 2018.


\bibitem{yaron}
Tomer Yaron, Avi Klein, Hamootal Duadi, and Moti Fridman, "Temporal superresolution based on a localization microscopy algorithm," Applied Optics, vol. 56, no. 9, p. D24, Mar. 2017.


\bibitem{Kolner2001}
C.V. Bennett and B.H. Kolner, "Aberrations in temporal imaging," IEEE Journal of Quantum Electronics, vol. 37, no. 1, pp. 20–32, Jan. 2001.




\bibitem{Kolner89}
Brian H. Kolner and Moshe Nazarathy, "Temporal imaging with a time lens," Optics Letters, vol. 14, no. 12, p. 630, Jun. 1989.


\bibitem{Bennette2000}
Corey V Bennett, "Principles of Parametric Temporal Imaging—Part I: System Configurations," IEEE JOURNAL OF QUANTUM ELECTRONICS, vol. 36, no. 4, p. 8, 2000.


\bibitem{Bennette20000}
Corey V Bennett, "Principles of Parametric Temporal Imaging—Part II: System Performance," IEEE JOURNAL OF QUANTUM ELECTRONICS, vol. 36, no. 6, p. 7, 2000.


\bibitem{Suxena}
Manish Saxena, Gangadhar Eluru, and Sai Siva Gorthi, "Structured illumination microscopy," Advances in Optics and Photonics, vol. 7, no. 2, p. 241, Jun. 2015.


\bibitem{Kaveh}
Ali Shayei, Zahra Kavehvash, and Mahdi Shabany, "Improved-resolution millimeter-wave imaging through structured illumination," Applied Optics, vol. 56, no. 15, p. 4454, May 2017.


\bibitem{wang18}
C. Wang, M. Zhang, B. Stern, M. Lipson, and M. Lončar, "Nanophotonic lithium niobate electro-optic modulators," Opt. Express  26, 1547-1555 (2018).


\bibitem{LJW}
Wright, L. J. Quantum pulse shaping by direct temporal phase modulation (Ph.D. dissertation), St Cross College, Oxford.

\bibitem{mishali}
M. Mishali and Y. C. Eldar, "From Theory to Practice: Sub-Nyquist Sampling of Sparse Wideband Analog Signals," IEEE Journal of Selected Topics in Signal Processing, vol. 4, no. 2, pp. 375–391, Apr. 2010.
\bibitem{BoLi}
Bo Li and Shuqin Lou, "Elimination of Aberrations Due to High-Order Terms in Systems Based on Linear Time Lenses," J. Lightwave Technol. 31, 2200-2206 (2013)
\bibitem{Pengyu}
Pengyu Guan, Kasper Meldgaard Røge, Mads Lillieholm, Michael Galili, Hao Hu, Toshio Morioka, and Leif Katsuo Oxenløwe, "Time Lens-Based Optical Fourier Transformation for All-Optical Signal Processing of Spectrally-Efficient Data," J. Lightwave Technol. 35, 799-806 (2017)

\bibitem{Zhao}
Zhao Wu, Lei Lei, Jianji Dong, Jie Hou, and Xinliang Zhang, "Reconfigurable Temporal Fourier Transformation and Temporal Imaging," J. Lightwave Technol. 32, 3963-3968 (2014)

\bibitem{saleh}
B. E. A. Saleh and M. C. Tiech, "Ultrafast optics," in "Fundamentals of
Photonics", (2007), pp. 936–1015.

\bibitem{500GIG1}
M. Burla et al., “500 GHz plasmonic Mach-Zehnder modulator enabling
sub-THz microwave photonics,” APL Photonics, vol. 4, no. 5, p. 056106,
May 2019

\bibitem{500GIG2}
C. Wang, M. Zhang, B. Stern, M. Lipson, and M. Loncar, "Nanophotonic 
lithium niobate electro-optic modulators," Opt. Express 26, 1547-1555
(2018).

\bibitem{zhan}
X. Zhao, S. Xiao, C. Gong, T. Yi, and S. Liu, “The implementation of temporal synthetic aperture imaging for ultrafast optical processing,” Optics Communications, vol. 405, pp. 368–371, Dec. 2017.


\end{thebibliography}
\end{document}